\documentstyle[12pt]{article}

\begin{document}
\pagestyle{empty}
\begin{flushright}
{BROWN-HET-977} \\
{DAMTP 94-102} \\
{December 1994}
\end{flushright}
\vspace*{5mm}
\begin{center}
{\bf PARTICLE PHYSICS MODELS, TOPOLOGICAL DEFECTS \\AND\\ ELECTROWEAK
BARYOGENESIS} \\
[10mm]
\renewcommand{\thefootnote}{\alph{footnote}}
Mark Trodden$^{1,2,}$\footnote{e-mail: mtrodden@het.brown.edu.},
Anne-Christine Davis$^{1,3,}$\footnote{e-mail:
A.C.Davis@damtp.cambridge.ac.uk.},
Robert Brandenberger$^{1,2,}$\footnote{e-mail: rhb@het.brown.edu.}\\
[10mm]

\end{center}
\begin{flushleft}
1){\it Isaac Newton Institute for Mathematical Sciences \\
University of Cambridge, Cambridge, CB3 OEH, U.K.}\\
2){\it Physics Department, Brown University,Providence RI. 02912. USA.} \\
3){\it Department of Applied Mathematics and Theoretical Physics and Kings
College, University of Cambridge, Cambridge CB3 9EW. U.K.}\\
\end{flushleft}
\begin{center}
{\bf Abstract}
\end{center}
\vspace*{3mm}
We demonstrate the viability of electroweak baryogenesis scenarios in which the
necessary departure from equilibrium is realized by the evolution of a network
of topological defects. We consider several effective models of TeV physics,
each addressing a fundamental particle physics problem, and in which the
conditions necessary for defect-mediated electroweak baryogenesis are naturally
satisfied. In each case we compare the strength of the model with that expected
from scenarios in which baryogenesis proceeds with the propagation of critical
bubbles.

\setlength{\textheight}{8.5in}
\newpage\setcounter{page}{1}\pagestyle{plain}
\renewcommand{\thefootnote}{\arabic{footnote}}
\section{Introduction}
The past few years have seen a great deal of effort devoted to attempts to
explain the generation of the baryon asymmetry of the universe (BAU) at the
electroweak scale\cite{TZ 90}-\cite{JPT2 94} (for reviews see \cite{{Turok
Review},{CKNi 93}}).

Within the context of the Weinberg-Salam theory of electroweak interactions it
is possible to satisfy all three of the Sakharov\cite{sakharov} criteria
necessary for a particle physics model to generate a net baryonic excess. These
criteria are
\begin{enumerate}
\item the existence of baryon number violating processes,
\item C and CP violation,
\item departure from thermal equilibrium.
\end{enumerate}
Common to all scenarios is the use of finite temperature sphaleron transitions
to achieve the first of these\cite{NM 83}-\cite{KRS 85}. Also, most authors
invoke an extended Higgs sector to obtain sufficient CP violation (C is
violated maximally) in the model (see \cite{FS 93} and \cite{NT 94} for
attempts to relax this). Recent attention has focussed on two alternative ways
to achieve the departure from equilibrium which is also required.

The most common scenario for weak scale baryogenesis involves a strongly first
order electroweak phase transition which is assumed to proceed by the
nucleation and propagation of critical bubbles of the true vacuum in the false.
It is in the walls of these bubbles that the changing Higgs fields communicate
the departure from equilibrium to the other particle species\cite{TZ
90}-\cite{JPT2 94}.

However, there exists an alternative realization of the third Sakharov
criterion in the context of the electroweak phase transition in the presence of
topological defects remaining after a previous symmetry breaking\cite{BDT 94}.
This is made possible by the fact that the electroweak symmetry may be restored
out to some distance around these defects and thus the evolution of the defect
network provides a departure from equilibrium in an analogous manner to that
produced by bubble walls.

In previous papers\cite{BDT 94} we have analyzed the details of this mechanism
in a general context without reference to a specific particle physics
implementation. We have examined general symmetry breaking schemes

\begin{equation}
G \stackrel{\eta}{\longrightarrow} SU(3)_c \times SU(2)_L \times U(1)_Y
\stackrel{\eta_{EW}}{\longrightarrow} SU(3)_c \times U(1)_{em}
\end{equation}
such that the first stage produces topological defects of a given type
determined by the non-triviality of the appropriate homotopy group of the
vacuum manifold ${\cal M}\equiv G/SU(3)\times SU(2) \times U(1)$. If
$\pi_1({\cal M}) \neq 1$ we obtain cosmic strings and if $\pi_0({\cal M}) \neq
1$ we have domain walls. In each case we have examined the baryon to entropy
ratio expected to be produced by the evolution of the defect network.

In this letter our aim is to outline examples of existing effective TeV
theories in which the criteria set forth in our earlier work are satisfied.
Thus we shall demonstrate the viability of defect-mediated electroweak
baryogenesis in the context of existing particle physics models.

It is interesting to note that the original motivations for studying the models
we describe are the resolution of particular particle physics problems. It is
therefore satisfying that the structure we require is an existing feature of
the models.

The outline of this letter is as follows. In section 2 we shall give a brief
summary of defect-mediated baryogenesis and define the relevant model-dependent
quantities. In section 3 we consider baryogenesis in the {\it Aspon model}\
\cite{aspon} and a supersymmetric (SUSY) model with an extra $U(1)$
symmetry\cite{susy}. Section four describes how the scenario is realized in a
model with a discrete family symmetry\cite{family} and in section 5 we
conclude.

\section{Defect-Mediated Electroweak Baryogenesis}
It has recently been shown that topological defects produced at scales above
the weak scale may restore the electroweak symmetry out to some distance $R_s$
around their core\cite{PD 93}-\cite{MT 94}. Since the electroweak Higgs vacuum
expectation value (VEV) is zero in this region we expect that just after the
weak phase transition baryon number violating processes will be unsuppressed in
the symmetric phase.

Shortly after the electroweak phase transition the evolution of the defect
network leads to a departure from thermal equilibrium in the walls of the
defects in the same manner as the motion of phase boundaries in other
mechanisms. Sufficient CP violation also occurs in the walls. This is assumed,
as in other models to come from an extended Higgs sector. Thus all three
Sakharov conditions are satisfied.

The final baryon to entropy ratio produced by such a scenario may be written as

\begin{equation}
\frac{n_b}{s}=\frac{n^{(0)}_b}{s}(1-e^{-Q})({\rm SF})
\end{equation}
where $n^{(0)}_b/s$ is the baryon to entropy ratio produced by a comparable
bubble wall mechanism, $Q$ is a factor connected with the competing effects
from different sides of the defect and (SF) given by

\begin{equation}
({\rm SF}) = \left(\frac{V_{BG}}{V}\right)
\end{equation}
is the volume suppression factor. This is the fraction of space swept out by
the defect network.

The factor $Q$ is dependent only on the type of baryogenesis considered: local,
where baryon number violation and CP violation take place at the same spacetime
point or nonlocal, where the two act in different regions; CP violation leads
to asymmetries in quantum numbers othere than baryon number on the wall and
these are then converted into baryon number in the larger region of symmetry
restoration. We shall comment briefly on this later. However, (SF) is dependent
on the geometry of the defects formed and their evolution. This must be
evaluated separately in each model.

\section{Models with TeV Scale Cosmic Strings}
The particle physics literature contains many examples which admit linear
topological defects - cosmic strings - some occurring at or around the TeV
scale. Here we shall give two examples and an estimate of the baryon to entropy
ratio which they may produce.

\subsection{The Aspon Model}
The Aspon model\cite{aspon} is intended as a resolution of the strong CP
problem of the standard strong and electroweak theory. In Quantum
Chromodynamics (QCD) instanton effects induce CP violating interactions. These
effects contribute to the electric dipole moment of the neutron by an amount
which disagrees with experiment unless the dimensionless parameter $\theta$
which measures their strength is kept small or zero.

The Aspon model achieves this by extending the gauge group of the standard
model by a new $U(1)$ symmetry. This leads to CP being a symmetry of the
Lagrangian. The $U(1)_{new}$ symmetry is then required to be broken at a scale
$\eta$. Thus the symmetry breaking scheme is

\begin{equation}
 SU(3)_c \times SU(2)_L \times U(1)_Y \times U(1)_{new}
\stackrel{\eta}{\longrightarrow} SU(3)_c \times SU(2)_L \times U(1)_Y
\stackrel{\eta_{EW}}{\longrightarrow} SU(3)_c \times U(1)_{em}
\end{equation}
(c.f. equation 1).

In addition to the particle content of the standard model (with a two-Higgs
structure for extra CP violation in the Higgs sector) the simplest example
contains two vectorlike quarks and two Higgs scalars $\chi_1$, $\chi_2$,
singlets under the standard model gauge group, which break the $U(1)_{new}$.
Two such fields are required so that the phase $\theta$ which is adjusted to
solve the strong CP problem cannot be rotated away by a gauge transformation.

It is assumed that these scalars get VEV's

\begin{equation}
\langle\chi_1 \rangle=\frac{1}{\sqrt{2}}\kappa_1 e^{i\theta}\ \ \ \ \ \, \ \ \
\ \ \langle\chi_2 \rangle=\frac{1}{\sqrt{2}}\kappa_2
\end{equation}
In order to obtain a simultaneous fit to weak and strong CP phenomenology it is
required that

\begin{equation}
\kappa < 2\  {\rm TeV}
\end{equation}
where $\kappa^2 \equiv \kappa_1^2 +\kappa_2^2$. Thus the extra symmetry
$U(1)_{new}$ is broken at a scale $\eta\sim~1$TeV and cosmic strings are
produced. After the electroweak phase transition the electroweak symmetry
remains restored around these objects. Note that since the extra sector is
fitted to weak CP data we expect to need an additional source of CP violation
in the standard model Higgs sector to implement baryogenesis.

\subsection{Supersymmetry with an Extra $U(1)$}
The particular supersymmetric model we consider\cite{susy} is proposed as a
solution to the $\mu$-problem of the minimal supersymmetric standard model
(MSSM) and the cosmological solar neutrino problem.

In the MSSM there exists a supersymmetric Higgs mixing term of the form

\begin{equation}
{\cal L}_{\mu} = \mu {\bar H}H
\end{equation}
In order to obtain radiative SUSY breaking at the weak scale it is necessary
that $\mu\sim {\cal O}(G_F^{-1/2})$ where $G_F$ is the Fermi constant. However,
there is no natural scale in the MSSM to ensure that this is the case.

In the model under consideration the MSSM is supplemented by two $U(1)$
symmetries. One of the extra $U(1)$'s breaks at a high scale ($\sim
10^{15}$GeV) and is concerned with the implementation of the MSW\cite{MSW}
solution of the solar neutrino problem via the seesaw mechanism. We shall not
discuss this further.
The $\mu$-term in this model is given in terms of a Yukawa coupling $\lambda'$
and a scalar $S$ which is a singlet under the standard model gauge group but
charged under the low energy extra $U(1)$. Thus the term (7) is forbidden and
in its place we have a term

\begin{equation}
{\cal L}_{\mu} = \lambda' S{\bar H}H
\end{equation}
Therefore if the low energy $U(1)$ breaks at a scale $\eta$ of the order of
1TeV then $S$ gets a VEV of this order and the $\mu$-problem is resolved.

Thus the symmetry breaking scheme of the model is

\begin{eqnarray}
SU(3)_c \times SU(2)_L \times U(1)_Y \times U(1) \times U(1) & \longrightarrow
&  SU(3)_c \times SU(2)_L \times U(1)_Y \times U(1) \nonumber \\
 & \stackrel{\eta}{\longrightarrow} &  SU(3)_c \times SU(2)_L \times U(1)_Y
\nonumber \\
 & \stackrel{\eta_{EW}}{\longrightarrow} &  SU(3)_c \times U(1)_{em}
\end{eqnarray}
Clearly we obtain TeV scale cosmic strings from this final $U(1)$ breaking.

\subsection{Cosmic Strings and Electroweak Baryogenesis}
Both models described above give rise to cosmic strings with a mass per unit
length of ${\cal O}(1TeV^2)$. Now, a string with mass per unit length $\mu$
remains in the friction dominated epoch until\cite{KEH} a time

\begin{equation}
t^*=(G\mu)^{-1}t_c
\end{equation}
where $t_c$ is the time of the defect-forming phase transition. For our strings
the corresponding temperature is $T^*=10^{-3}$TeV$=1$GeV$<\eta_{EW}$. Therefore
the string network is still in the friction dominated era at the electroweak
phase transition.

The contribution to the baryon asymmetry comes from both string loops and
``infinite" strings. Let us focus on long strings, since in the friction
dominated era most of the energy of the network is in this form. The
suppression factor for long strings can easily be shown to be (see \cite{BDT
94} for details)

\begin{equation}
(SF) = \lambda v_D\left(\frac{\eta_{EW}}{\eta}\right)^{3/2}
\end{equation}
where $\lambda$ is the standard model Higgs self-coupling and $v_D$ is the
velocity of the defect. If we assume one ``infinite" string per correlation
volume this yields
\begin{equation}
(SF) \sim \frac{\lambda v_D}{30}
\end{equation}
for $\eta \sim 1$TeV.

If we consider nonlocal baryogenesis then the final baryon to entropy ratio is
given by (2) with $n_b^{(0)}/s$ given by (see \cite{JPT3 94})

\begin{equation}
\frac{n_b^{(0)}}{s} \simeq 10^{-6}\kappa \Delta \theta y_{\tau}^2 v_D
\end{equation}
where $0.1 \leq \kappa \leq 1$ is defined by $\Gamma_B=\kappa(\alpha_W T)^4$
with $\Gamma_B$ the rate per unit volume of baryon number violating processes
in the region of unbroken electroweak symmetry\cite{{ALS 89},{GST 92}}. Here
$\Delta\theta$ is a measure of the CP violation and $y_{\tau}$ is the Yukawa
coupling of the $\tau$-lepton, the scattering of which we have focussed on.

Thus this scenario can be compatible with the required nucleosynthesis value of
$n_b/s \sim 10^{-10}$.

\section{Domain Walls from Family Symmetries}
We shall consider particle physics models which attempt to explain the mass
hierarchy in the standard model, in particular the large top quark mass, using
discrete family symmetries\cite{family}.

In general the symmetry breaking scheme of such a model is

\begin{eqnarray}
SU(3)_c \times SU(2)_L \times U(1)_Y \times G & \longrightarrow &  SU(3)_c
\times SU(2)_L \times U(1)_Y \times G^1 \nonumber \\
 & \longrightarrow &  SU(3)_c \times SU(2)_L \times U(1)_Y \times G^2 \nonumber
\\
 & \vdots & \nonumber \\
 & \longrightarrow &  SU(3)_c \times SU(2)_L \times U(1)_Y \times G^n \nonumber
\\
 & \stackrel{\eta}{\longrightarrow} &  SU(3)_c \times SU(2)_L \times U(1)_Y
\nonumber \\
 & \stackrel{\eta_{EW}}{\longrightarrow} &  SU(3)_c \times U(1)_{em}
\end{eqnarray}
where $G \supset G^1 \supset G^2 \supset \cdots \supset G^n$ are nested
discrete groups, the symmetry breakings between which lead to the generation of
the mass hierarchy.

It is commonly assumed that these finite groups are gauged in the sense that G
arises from the spontaneous breaking of a continuous group $H$. This is
necessary to protect the theory from the wormhole effects thought to plague
global symmetries.

For definiteness let us concentrate on a specific example

\begin{equation}
H=SU(2),\ \ \ \ \ \ \ \ \ \ G=Q_6, \ \ \ \ \ \ \ \ \ \ G^n={\cal Z}_6
\end{equation}
where $Q_6$ is the double dihedral group of order 12. We shall be interested
only in the final discrete breaking in which ${\cal Z}_6$ is broken completely
giving tree level masses to the strange quark and the $\mu$-meson. This
breaking occurs at the ubiquitous scale $\eta \sim {\cal O}(1$TeV$)$ and
produces cosmological domain walls. It is an important caveat that we require
that there be a mechanism to dispose of these walls after the electroweak phase
transition so that they do not dominate the energy density of the universe.

Since we expect the electroweak symmetry to be restored around the TeV domain
walls produced by the above theory it is simple to estimate the final baryon
asymmetry expected to be produced by their evolution.

Consider the effect of ''infinite" domain walls. Then the suppression factor
(SF) is given by (see \cite{BDT 94})

\begin{equation}
(SF) \sim {\cal O}(1)v_D
\end{equation}
Thus, using (2) and estimates for $n_b^{(0)}/s$ which may be $10^{-8}$ or even
smaller, the final baryon to entropy ratio can be compatible with observations.

\section{Conclusions}
We have discussed several specific particle physics models in which the
criteria for defect-mediated electroweak baryogenesis are satisfied. Each model
is an effective TeV theory with its own particle physics virtues.

Clearly there are more models which fulfill the requirements and the above are
a small sample which we believe demonstrate the viability of the scenario.

\vspace{1cm}
\begin{center}
\bf Acknowledgements
\end{center}
\vspace{5mm}
R.B. and M.T. thank Paul Frampton for useful suggestions. We thank the Isaac
Newton Institute, Cambridge University for hospitality. This work was supported
in part by the US Department of Energy under Grant DE-FG0291ER40688, Task A, by
an NSF Collaborative Research Award NSF-INT-9022895 and by PPARC.

\end{document}